\documentclass[10pt, sigplan, nonacm]{acmart}

\usepackage{booktabs}  
\usepackage{balance}
\usepackage{xcolor}
\usepackage{multicol}
\usepackage{multirow}
\usepackage{multitoc}
\usepackage{textcomp}
\usepackage{xcolor}
\usepackage[ruled,vlined]{algorithm2e}
\usepackage{graphicx}
\usepackage{float}
\usepackage{algorithmic}
\usepackage{soul}
\usepackage{caption}
\usepackage{subcaption}
\usepackage{listings}
\usepackage{verbatim}
\usepackage{lstcustom}

\newcommand\ie[1]{e.g., }
\definecolor{blueCode}{RGB}{0, 0, 128}

\usepackage[normalem]{ulem}

\newcommand\acode[1]{%
    \textcolor{blueCode}{\texttt{#1}}
}

\newcommand\toolkit{%
    Beehive SPIR-V Toolkit
}

\AtBeginDocument{%
  \providecommand\BibTeX{{%
    \normalfont B\kern-0.5em{\scshape i\kern-0.25em b}\kern-0.8em\TeX}}}

\begin{document}



\title{Experiences in Building a Composable and Functional API for Runtime SPIR-V Code Generation}

\author{Juan Fumero}
\email{juan.fumero@manchester.ac.uk}
\affiliation{%
  \institution{University of Manchester}
  \streetaddress{Oxford Road}
  \city{Manchester}
  \country{UK}
  \postcode{M13 9PL}
}

\author{Gy\"orgy Rethy}
\authornote{Work done while he was associated at The University of Manchester.}
\email{grethy@student.ethz.ch}
\affiliation{%
  \institution{ETH Zurich}
  \streetaddress{XXX}
  \city{Zurich}
  \country{Switzerland}
  \postcode{XXXX}
}

\author{Athanasios Stratikopoulos}
\email{athanasios.stratikopoulos@manchester.ac.uk}
\affiliation{%
  \institution{University of Manchester}
  \streetaddress{Oxford Road}
  \city{Manchester}
  \country{UK}
  \postcode{M13 9PL}
}

\author{Nikos Foutris}
\email{nikos.foutris@manchester.ac.uk}
\affiliation{%
  \institution{University of Manchester}
  \streetaddress{Oxford Road}
  \city{Manchester}
  \country{UK}
  \postcode{M13 9PL}
}

\author{Christos Kotselidis}
\email{christos.kotselidis@manchester.ac.uk}
\affiliation{%
  \institution{University of Manchester}
  \streetaddress{Oxford Road}
  \city{Manchester}
  \country{UK}
  \postcode{M13 9PL}
}


\begin{abstract}


This paper presents the Beehive SPIR-V Toolkit; a framework that can automatically generate a Java composable and functional library for dynamically building SPIR-V binary modules.
The Beehive SPIR-V Toolkit can be used by optimizing compilers and runtime systems to generate and validate SPIR-V binary modules from managed runtime systems, such as the Java Virtual Machine (JVM).
Furthermore, our framework is architected to accommodate new SPIR-V releases in an easy-to-maintain manner, and it facilitates the automatic generation of Java libraries for other standards, besides SPIR-V.
The Beehive SPIR-V Toolkit also includes an assembler that emits SPIR-V binary modules from disassembled SPIR-V text files, and a disassembler that converts the SPIR-V binary code into a text file, and a console client application.
To the best of our knowledge, the Beehive SPIR-V Toolkit is the first Java programming framework that can dynamically generate SPIR-V binary modules.

To demonstrate the use of our framework, we showcase the integration of the SPIR-V Beehive Toolkit in the context of the TornadoVM, a Java framework for automatically offloading and running Java programs on heterogeneous hardware. 
We show that, via the SPIR-V Beehive Toolkit, the TornadoVM is able to compile code 3x faster than its existing OpenCL C JIT compiler, and it performs up to 1.52x faster than the existing OpenCL C backend in TornadoVM. 
\end{abstract}



\keywords{API, Library, Java, Metaprogramming, Runtime Code Generation, SPIR-V}

\maketitle

\section{Introduction}

The Standard Portable Intermediate Representation (SPIR-V)~\cite{spirv-standard}, maintained by the Khronos group~\cite{khronos}, is an intermediate binary format for representing graphics and parallel computation that exploit parallel execution on heterogeneous hardware, such as GPUs and FPGAs.
SPIR-V was proposed in March 2015~\cite{spirv-standard}, as an extension to OpenCL compute kernels for enabling applications to consume SPIR-V binaries instead of computing OpenCL C programs. 
Several companies, including Intel, AMD, NVIDIA, Xilinx, Codeplay and Google offer their own implementations, tools and compilers for generating and consuming SPIR-V binary code. 
Besides those tools, the Khronos Group has also created and maintained a set of tools and utilities to compile, disassemble, validate and optimize SPIR-V binary code~\cite{spirvtools}. 
However, all these tools are available for LLVM-based programming languages implementations, such as C/C++ (e.g., \texttt{clang}~\cite{lattner2008llvm}), as they operate under the LLVM ~\cite{10.5555/977395.977673} ecosystem and compiler infrastructure\footnote{
Note that, although SPIR-V was created to be independent of the LLVM-IR, many of the tools and utilities are still under the LLVM and C++ compiler infrastructure.}.

This feature hinders the exploitation of SPIR-V tools from programming languages, such as Java, R, Ruby, and Scala, which are built on top of managed runtime systems, such as the Java Virtual Machine (JVM).
Applications and systems' software, such as optimizing compilers and runtime systems (e.g., GraalVM~\cite{duboscq2013graal}), that are implemented in those programming languages, cannot use existing standard SPIR-V tools in a direct way.

To enable the utilization of LLVM-based tools from managed programming languages, it is necessary to invoke the tools using native interfaces (e.g.,\ Java Native Interface (JNI)).
For instance, Java could invoke existing LLVM tools by providing native methods that can be used as library calls via JNI.

Another way is to invoke LLVM utilities as an invocation of a new process from a guest programming language (e.g., by creating a subprocess that invokes the LLVM tools from Java).
However, such integrations pose the following challenges: i) high complexity due to the interaction between different programming languages and the runtime system, ii) high difficulty to maintain, and iii) the necessity to recompile the Java native code for every new release of SPIR-V.

In this paper, we present the Beehive SPIR-V Toolkit, a Java framework that enables the automatic generation of a composable and functional Java library based on standardized grammar files (e.g.,\ SPIR-V grammar).
The generated library can be used to enhance software written in JVM programming languages with functionality (i.e.,\ generation, validation, optimization) that is currently offered by LLVM-based SPIR-V tools.
Besides, the proposed framework is architected to accommodate new SPIR-V releases in an easy-to-maintain manner and facilitate the automatic generation of Java libraries for other standards, besides SPIR-V.

To enable this functionality, the Beehive SPIR-V Toolkit architecture encompasses a template system engine that demonstrates a well-known technique based on model-driven engineering~\cite{favre2004towards} for automatically generating Java libraries and composable APIs based on standardized specifications.
Note that this is a common software engineering technique, and we have enabled it to provide new APIs that can be used by compilers and runtime systems in a transparent manner without the need to reconfigure new rules for every new version of the SPIR-V standard. 
Besides, this technique can be used by system architects to build new libraries for other standards.

Lastly, the \toolkit provides a client utility that can: i) assemble a SPIR-V binary code from a kernel description that is stored in a text file, and ii) disassemble a binary code to a kernel description stored in a file.

The \toolkit has been developed to be utilized by existing Just-In-Time (JIT) compilers and managed runtime systems, as a means to facilitate SPIR-V code generation and analysis from high-level JVM programming languages.
The source code of the Beehive SPIR-V Toolkit is available on GitHub\footnote{\url{https://github.com/beehive-lab/beehive-spirv-toolkit}} as an open-source project under the MIT license.
As a use case, we have extended the TornadoVM~\cite{10.1145/3313808.3313819, 10.1145/3237009.3237016} (a heterogeneous programming framework for Java) JIT Compiler and runtime system, with a new backend for generating SPIR-V using the \toolkit as a library.

In a nutshell, this paper makes the following contributions:
\begin{itemize}
    \item It presents a template-based technique to automatically generate composable and functional Java programming libraries from standard grammar files that are defined using the JSON format, showcasing the technique in the context of SPIR-V.
    \item It presents the Beehive SPIR-V Toolkit, a framework that automatically generates a composable and functional Java library for building SPIR-V binary modules at runtime.  
    The proposed framework includes an assembler that emits SPIR-V binary modules from disassembled SPIR-V text files, as well as a disassembler that converts the SPIR-V binary code to a kernel description stored in a text file.
    \item It presents an extension of the TornadoVM JIT compiler and runtime system for automatically generating SPIR-V from Java bytecode. 
    \item  It presents a performance evaluation of our framework against existing OpenCL C backend of TornadoVM, showing end-to-end speedups of up to 3x for the code generation and up to 1.52x speedup for the execution of the generated code.
\end{itemize}

\section{Background and Motivation}

SPIR-V was created to address the need for a universal intermediate language for parallel computing  and graphics processing applications. 
Similarly to OpenCL, SPIR-V allows developers to write their code once and then compile it to different target architectures, such as CPUs, GPUs, and FPGAs, among others, but in binary format. 
However, with SPIR-V, developers do not need to distribute the source code of the compute kernels, but rather the binary representation of those kernels. 
Additionally, SPIR-V supports extensions from multiple vendors and parties (e.g., SPIR-V Math Extended Instruction Set~\cite{spirvmath}).

Since SPIR-V is a binary format representation, it might be more convenient for compilers to generate this intermediate representation rather than, for example, OpenCL C source code. 
In addition, SPIR-V kernels can be consumed by OpenCL programs (from OpenCL 2.1) and Intel Level Zero~\cite{levelZero} applications. 

SPIR-V allows high-level language front-ends to produce programs in a standardized intermediate format that can be consumed by drivers for Vulkan~\cite{vulkan}, OpenGL~\cite{opengl} or OpenCL~\cite{stone2010opencl}, thus eliminating the need for high-level language front-end compilers in device drivers. 
This not only simplifies driver architecture but also supports a broad range of language and framework front-ends for different hardware architectures, promoting an active ecosystem of open-source tools for analysis, porting, debugging, and optimization.

Lastly, by using SPIR-V, developers can avoid exposing kernel source code, improve kernel load times (since it is in binary format), and use a common language front-end compiler to enhance kernel reliability and portability across multiple hardware implementations.

\begin{figure}[t]
    \includegraphics[width=0.45\textwidth]{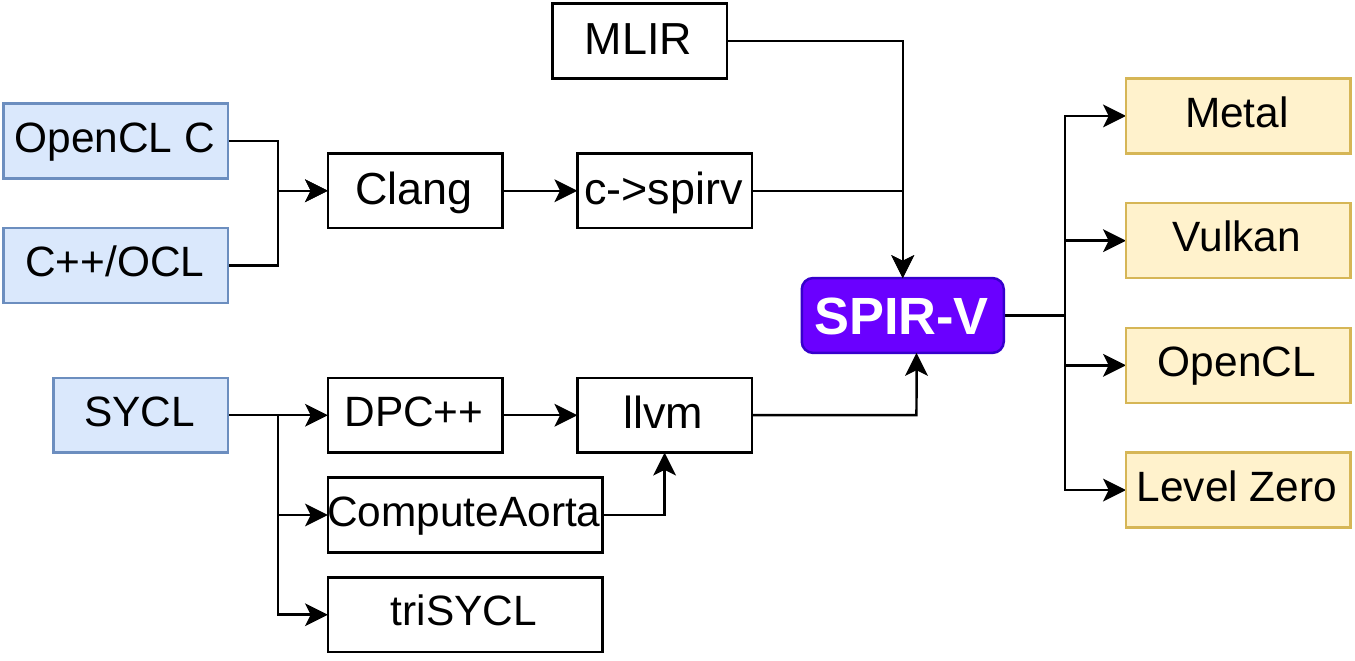}
    \caption{Language description and tooling ecosystem for SPIR-V as described in the SPIR-V SPEC~\cite{spirv-standard}.}
    \label{fig::spirvSPECFlow}
\end{figure}

\paragraph{But, what is missing?} 
Figure~\ref{fig::spirvSPECFlow} shows the SPIR-V Language ecosystem and some of the tools associated with SPIR-V as they are specified in the SPIR-V Standard~\cite{spirv-standard}.
From the figure, we can see that SPIR-V represents the intermediate step between the high-level languages and the low-level programming models for computing and graphics processing. 
From the high-level language descriptions, we see that all tools and languages represent conversions from the C and C++ programming languages, with a heavy focus on the LLVM compiler ecosystem. 

While SPIR-V was designed to be independent of the LLVM software and compiler ecosystem, current tools and translators are centered around LLVM. 
This paper presents a framework to generate SPIR-V code from managed runtime programming languages and managed runtime environments, such as the Java Virtual Machine and the Java programming language. 
As far as we know, this paper presents the first Java library to dynamically generate and validate SPIR-V code.

\begin{figure*}[t!]
    \centering    
    \includegraphics[width=0.83\textwidth]{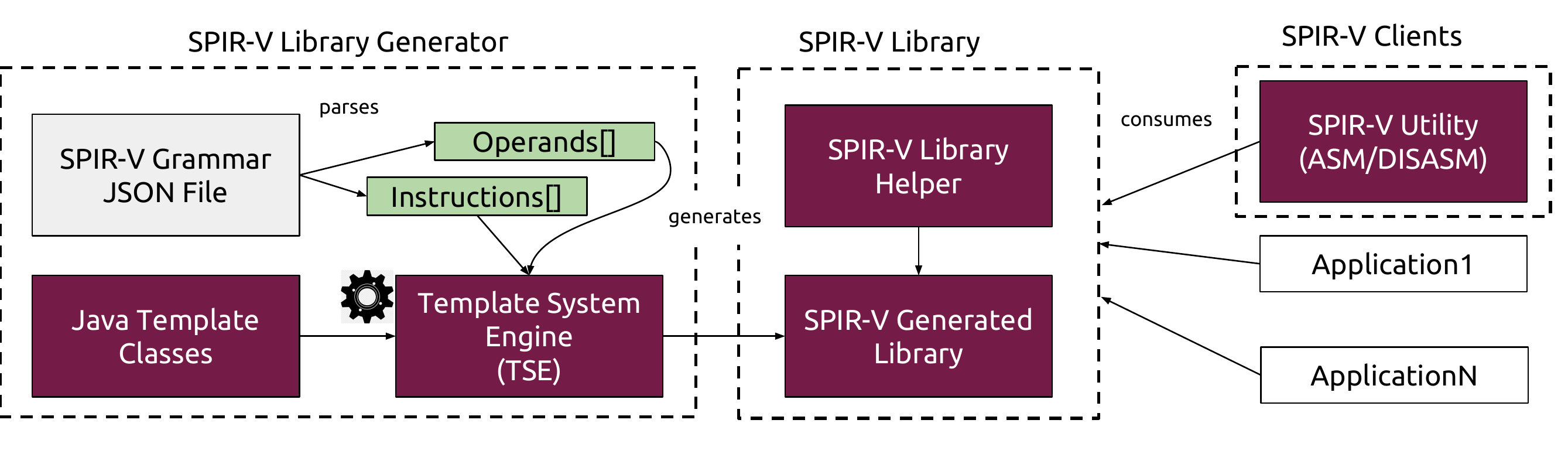}
    \caption{Overview of the main components of the Beehive SPIR-V Toolkit. 
    The Beehive SPIR-V Toolkit provides three main components: 1) the SPIR-V Library Generator; 2) the SPIR-V Library, and 3) a SPIR-V Client. 
    The purple sub-components are provided by the Beehive SPIR-V Toolkit. The SPIR-V Grammar file is provided by the Khronos Group in GitHub~\cite{spirvtools}.
    The green components represents two sub-components within the Library Generator that stores all SPIR-V instructions and operands that will be used to generate the new Java types.}
    \label{fig::overview}
\end{figure*}

\section{Beehive SPIR-V Toolkit}
\label{section::sbt}

This section presents the \toolkit.
First, it shows an overview of the overall software architecture, and it shows how the Beehive SPIR-V Toolkit works (Section~\ref{sub::overview})
Then, Sections~\ref{sub::generator}-\ref{sub:client} describe each component in detail. 
  
\subsection{System Overview}
\label{sub::overview}

The \toolkit is a Java framework that generates a Java library and a Java API for composing SPIR-V modules\footnote{A SPIR-V module is a single compilation unit that can contain multiple entry points that share the same capabilities (e.g., compute capabilities).}.
The generated Java library is designed to be used as an assembler and disassembler of SPIR-V binary modules at runtime by managed runtime systems.

To generate the Java library and the API, we have implemented a Template System Engine (TSE) (Section~\ref{sec::templateSystem}).
The TSE is also fully implemented in Java, and it generates new Java types that will be used to compose SPIR-V binary modules. 
The generated library is then distributed as a standard Java JAR file to be imported by Java client applications, runtime systems and compilers that are also implemented in Java.
Figure~\ref{fig::overview} shows the three main components, as follows:

\paragraph{SPIR-V Library Generator:} This component generates the Java SPIR-V Library as well as the APIs for composing SPIR-V binary modules.
The SPIR-V Library Generator takes, as inputs, a set of grammar files described in the JSON format that represents the SPIR-V grammar specified by the Khronos standard, and a set of templates implemented in Java.
The Java templates provide the initial scaffolding to facilitate the automatic generation of the functional and composable APIs, as well as helper Java classes.

The core of the template generator is the TSE component, which is a fully automated system that allows the generation of composable APIs for every new version of the SPIR-V standard based on the new grammar files. 
We refer to the generated API as \emph{composable} because, in order for client applications to build new SPIR-V modules, it makes use of function composition to create and build new constructs for the resulting SPIR-V binary module. 
Section~\ref{sub::library} will show several examples of type and construct composition. 

\paragraph{SPIR-V Library:} This component is the result of the SPIR-V library generator, and it is meant to be distributed to Java clients as a standard Java library in a JAR format. 
This library allows developers to implement JVM-based programs (e.g., those implemented in Java, Scala, etc.) that dynamically create SPIR-V modules.
Furthermore, the SPIR-V library includes functionality to check and validate basic rules for SPIR-V modules. 

\paragraph{SPIR-V Client Utility:} This component provides a client application that can be used for assembling and disassembling SPIR-V modules as a standalone tool.
The client application assembles and disassembles SPIR-V code from text to binary format and vice-versa.  
Furthermore, this component acts as an end-to-end application that demonstrates how the generated SPIR-V library can be used with complete examples. 


\subsection{SPIR-V Template Library Generator}
\label{sub::generator}

The objective of the SPIR-V template library generator is to automatically generate Java libraries that can be utilized by other software components (e.g.,\ optimizing compilers, runtime systems, etc.) as a means to develop SPIR-V code.
The generated API is \textbf{functional} (as in functional programming in which each new operation returns a new object of the requested type, such as an integer addition, and there is no mutation state of internal properties of the Java objects), and \textbf{composable} (new instructions are composed of other instructions, that are, in turn, built by creating and composing new objects of the requested types).
We demonstrate the details of the composition of calls in Section~\ref{sub::library}. 

Furthermore, the library generator is capable of creating an assembler and a disassembler that can adhere to various versions of the SPIR-V standards.
Thus, any future extensions in the SPIR-V standard can be easily adopted in existing systems' software by updating to newly generated SPIR-V formats in a transparent and automatic manner.
To enable this functionality, the SPIR-V library generator employs two open-source software components: (1) the Jackson Annotation Library~\cite{jacksonAnnotationFramework}, to parse the JSON grammar files and create a list of instructions and operands to be generated; and (2) the Apache Freemarker~\cite{apacheFreeMarker}, to generate all Java types for all instructions and operands that were previously parsed.

\begin{figure}[t!]
    \begin{lstlisting}[caption=First Block Level of JSON Parsing annotated using the Jackson Annotation Framework., label={code:sample1}, xleftmargin=.03\textwidth, language=Java] 
@JsonIgnoreProperties(ignoreUnknown = true)
public class SPIRVGrammar {
  @JsonProperty("magic_number")
  public String magicNumber;
    
  @JsonProperty("instructions")
  public SPIRVInstruction[] instructions;
  
  @JsonProperty("operand_kinds")
  public SPIRVOperandKind[] operandKinds;
  ... 
}
\end{lstlisting}
\end{figure}

\begin{figure*}[t]
    \centering    
    \includegraphics[width=0.76\textwidth]{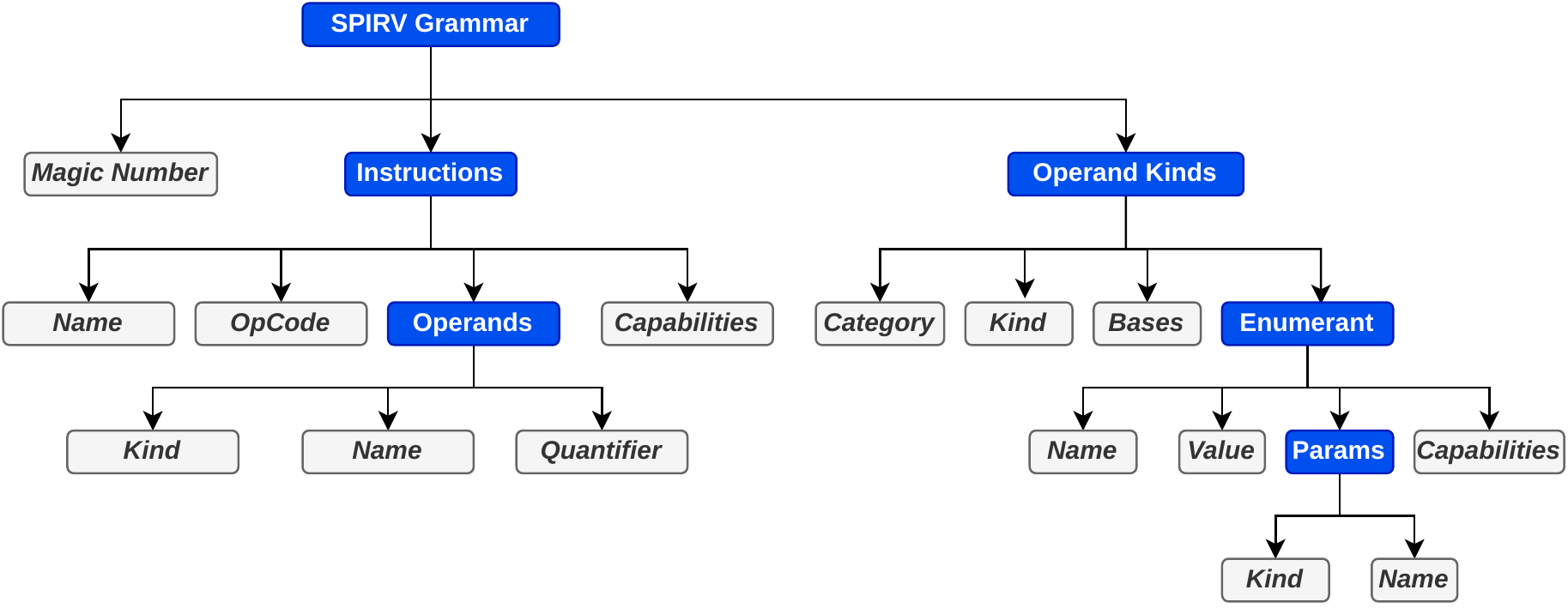}
    \caption{Block categories of the JSON file format that specifies the grammar of the SPIR-V Intermediate Language.}
    \label{fig::jsonParserSPIR-VGrammar}
\end{figure*}

\subsubsection{Parsing the JSON files from the SPIR-V Standard Specification} 
\label{subsub::parsing}
As part of the tools for the SPIR-V standard, the Khronos Group provides, in their repository\footnote{\url{https://github.com/KhronosGroup/SPIRV-Headers}}, the JSON files that specify the grammar for the SPIR-V Intermediate Language, and it defines all SPIR-V instructions, SPIR-V operands, and SPIR-V types. 

Figure~\ref{fig::jsonParserSPIR-VGrammar} presents the main structure of serialized JSON objects that reside in the JSON file. 
The blue blocks represent JSON objects that can be expanded to other SPIR-V category types, while the light-grey blocks represent final objects that correspond to SPIR-V final types.
As illustrated in Figure~\ref{fig::jsonParserSPIR-VGrammar}, the SPIR-V grammar is composed of a \textit{Magic Number} along with a set of \textit{Instructions} and a list of \textit{Operand Kinds}.

To parse the SPIR-V grammar JSON file and generate the corresponding Java classes that will compose the SPIR-V library, we use the Java Jackson annotation framework, which is a popular and widely used Java framework to map Java objects into/from files. 
Thus, the mapped Java objects from the JSON file are structured following the object definition in the standardized SPIR-V grammar. 

Listing~\ref{code:sample1} shows a sketch of Java code from the \toolkit that parses the first level of the SPIR-V grammar file. 
Each field within the Java class is annotated with \acode{@JsonProperty} (lines 3, 6, 9). 
This annotation contains a name with the exact field name as it is named in the JSON file. 
For instance, the field marked as \texttt{instructions} is represented in Java by an array of the \texttt{SPIRVInstruction} object (line 7).
Analogously, the \texttt{operand\_kinds} field is mapped to an array of \texttt{SPIRVOperandKind} Java objects (line 10). 

Furthermore, our library supports the OpenCL Extended Math Instructions set~\cite{spirvmath}.
To accommodate this extension, we introduced a set of Java classes annotated with Jackson annotations, similar to the parsing of the SPIR-V JSON file described in Listing~\ref{code:sample1}.
Note that, although the Beehive SPIR-V Toolkit implementation covers the whole SPIR-V grammar as defined in the SPIR-V specification, it does not cover the complete list of SPIR-V extensions, such as GLSL, and AMD extensions.
The reason is that at the current status of the \toolkit, our focus has been on providing support for parallel compute kernels (to be integrated with OpenCL and Intel Level Zero~\cite{levelZero} runtime systems) rather than graphics processing.
However, the proposed framework can be extended in the same way to generate all Java classes for graphics processing as well.

The SPIR-V Library Generator produces two kinds of Java classes.
The main two generated kinds are for representing: i) SPIR-V instructions; and ii) SPIR-V operands.
Additionally, the SPIR-V Library Generator produces mapper Java classes that can relate from text to SPIR-V binary format and vice-versa. 
These proxy Java classes are invoked when running the assembler and disassembler as standalone tools.
As follows, we will explain the format of the instructions and operand kinds as derived from the JSON files and how they are mapped to the generated SPIR-V library.

\paragraph{Parsing Instructions}
As shown in Listing~\ref{code:sample1}, SPIR-V instructions are represented as an array of \texttt{SPIRVInstruction} Java objects.
Every \texttt{SPIRVInstruction} object within that array defines a SPIR-V instruction from the SPIR-V specification. 
Each instruction is composed of a name (in a string format), an opcode, a list of operands, and a capability (e.g., \texttt{Shader} for graphics, or \texttt{Kernel} for expressing computation).
The field \texttt{Operands}, expands to a list of Java objects following the order that the list is specified in the grammar file.
For instance, an operand contains information about the kind of that operand, the name of the instruction, and a quantifier that describes the expected quantity of the operand (either optional or variadic).
To represent the quantifier of an operand, the following types are also included in the generated library: a) an optional operand, represented by the \texttt{SPIRVOptionalOperand} Java class; and b) multiple operands, represented by the \texttt{SPIRVMultipleOperands} class.
Note that these two Java types are not generated, but provided as basic types in the SPIR-V library, as we will discuss in Section~\ref{sec::templateSystem}.

In order to organize all SPIR-V instructions from the specification, we provide a root Java class called \texttt{SPIRVInstruction}.   
The \texttt{SPIRVInstruction} Java class is an abstract class that contains a set of methods to a) write instructions to a file in binary format, and b) return their properties via \textit{getter} methods (e.g., the opcode).
Most of the Java classes that implement SPIR-V instructions extend this abstract class.
For instance, the \texttt{SPIRVOpIAdd} type is a generated Java class that maps to the \texttt{OpIAdd} SPIR-V instruction and inherits from the \texttt{SPIRVInstruction} Java base class.  

\begin{figure}[t!]
    \centering    
    \includegraphics[width=1\columnwidth]{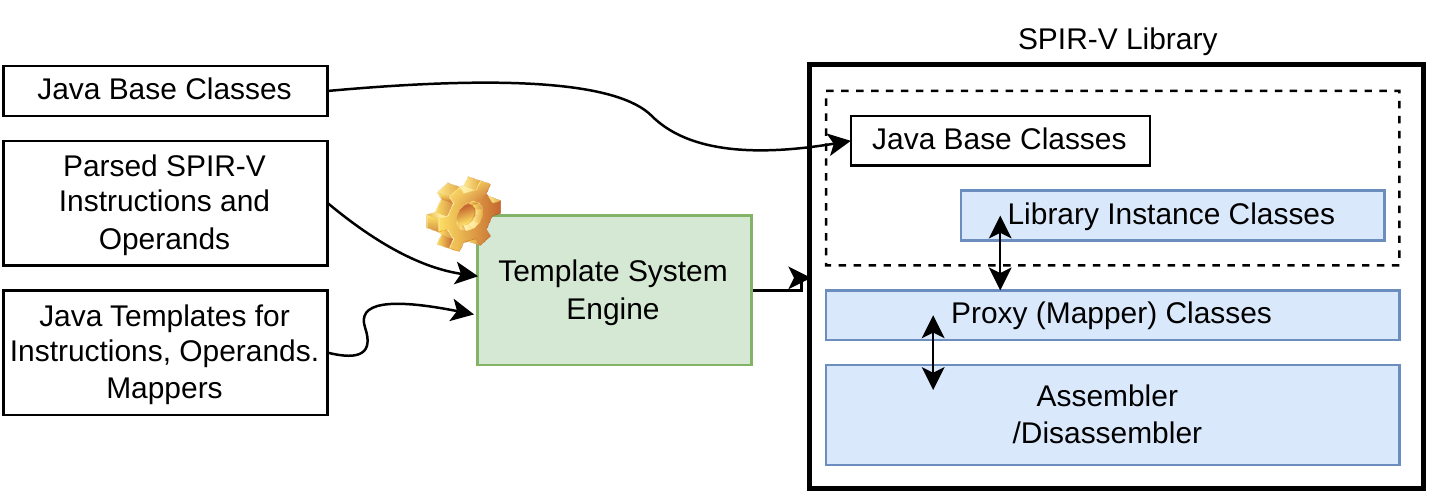}
    \caption{Overview of the workflow of the Template System Engine that generates the Beehive SPIR-V Toolkit Library.}
    \label{fig::templateEngine}
\end{figure}

\paragraph{Parsing Operand Kinds}
\label{subsec::operand_kinds}

An additional parsed JSON object from the SPIR-V grammar JSON file (as shown in Figure~\ref{fig::jsonParserSPIR-VGrammar}) is the \textit{Operand Kinds}. 
This field is represented by a Java object that describes a SPIR-V data type.
The data types defined in SPIR-V are classified as: \textit{Id (identifier)},  \textit{BitEnum (enumerate)}, \textit{ValueEnum}, \textit{Literal} and \textit{Composite}.
Each SPIR-V data type includes a category and the name of the type. 
Additionally, some operands may require specific capabilities from the hardware on which they will be compiled to execute. 

The Java classes that are generated for the \textit{Operand Kinds} follow the same structure (layout), as declared in the SPIR-V grammar files. 
Similar to the instruction's type, we have implemented a Java interface called \texttt{SPIRVOperand} as the root of the Java class hierarchy for all operands kinds.  
This interface declares three abstract methods for writing the operands into a binary format, getting the capabilities of an operand, and obtaining the word count (number of bytes that each instruction takes, thus, it will be easier to calculate the final amount of bytes for the resulting SPIR-V binary module).
These methods declare the basic functionality for the generated assembler and disassembler. 

Additionally, we have provided base Java classes for each data type as described in the SPIR-V specification. 
For instance, we have added Java classes for the \texttt{SPIRVEnum}, \texttt{SPIRVId}, \texttt{SPIRVLiteralContextDependentNumber}, \texttt{SPIRVString}, and \texttt{SPIRVMultipleOperands}.
These Java classes are included in the generated SPIR-V Library.
As we will discuss in the next section, these base Java classes are extended for every generated data type based on the SPIR-V specification. 
For instance, the generated class \texttt{SPIRVAddress-} \texttt{ingModel}, which indicates whether a SPIR-V kernel uses \texttt{logical}, \texttt{physical32} or \texttt{physical64} addressing mode, inherits from the \texttt{SPIRVEnum} Java base class. 

\begin{figure}[t]
    \begin{lstlisting}[caption=Pseudocode for the Instruction Generator., label={code:instructionGenerator}, xleftmargin=.04\textwidth, language=Java] 
Template t = openTemplate("instruction.template");
  forEach (i: grammar.getInstructions()) 
  do
    templateFile = createFile(i)    
    t.eprocess(templateFile, i);    
    t.writeFile();
  done
\end{lstlisting}
\end{figure}

\subsection{Template System Engine (TSE)}
\label{sec::templateSystem}

The core functionality of the SPIR-V Library Generator is the automatic generation of libraries through the Template System Engine (TSE), which generates a Java library that uses function composition to build, at runtime, SPIR-V binary modules. 

Figure~\ref{fig::templateEngine} shows a high-level overview of TSE component. 
The engine takes, as inputs, the Java classes with the parsed JSON objects, and the Java template classes.
The former classes map JSON objects to Java objects by the Jackson Annotation Framework (Section ~\ref{subsub::parsing}), while the latter Java classes are used for the template processing that will be used for generating all instructions, operands, and mappers.
Note that this is an automatic process, and it builds all classes the first time that the library is built while accepting a grammar JSON file as input. 

\paragraph{Porting Different Standard Versions}
We initially ported the SPIR-V 1.2 standard, because all major hardware and driver vendors (e.g., Intel, and NVIDIA) use this version. 
However, we also ported to the latest standard available, 1.6, or unified. 
We experience a smooth transition and we only needed to accommodate one of the templates due to the addition of a number at the beginning of a SPIR-V instruction.
Since we map every instruction to a Java class, we cannot assign a class name that starts with a number. 
Thus, in this case, we start with the symbol underscore for the generated Java type\footnote{\url{https://github.com/beehive-lab/spirv-beehive-toolkit/commit/217208a2ef9b7a7ec4eb4d59e32c45a2344d8582}}.

\paragraph{Library Organization}
As discussed in Section~\ref{subsec::operand_kinds}, a generated library contains a set of Java base classes that include abstract classes and interfaces.
Figure~\ref{fig::templateEngine} shows how the TSE component organizes the Java classes for the resulting library. 
The base classes provided by the Library Generator are forwarded to the generated SPIR-V Library. 
In addition, the TSE component generates a set of Java classes, mapper classes and the utilities for the assembler and dissembler based on the generation of new types described by the Java templates for representing all SPIR-V types for instructions and operands. 
As such, these Java templates follow three different categories within the TES component: a) instructions templates; b) operand templates (including composites, literals and enumerates); and c) mappers templates. 

\begin{figure*}[t!]
\begin{lstlisting}[caption=Example of a Template class for representing SPIR-V Instructions., label={code:templateInstruction}, xleftmargin=.03\textwidth, language=Java] 
@Generated("beehive-lab.SPIR-Vbeehivetoolkit.generator")
    public class SPIR-V${name} extends ${superClass} {
      <#if operands??>
      <#list operands as operand>
      <#if operand.quantifier == '*'>
            public final SPIR-VMultipleOperands<SPIR-V${operand.kind}> ${operand.name};
      <#elseif operand.quantifier == '?'>
            public final SPIR-VOptionalOperand<SPIR-V${operand.kind}> ${operand.name};
      <#else>
            public final SPIR-V${operand.kind} ${operand.name};
      </#if>
      </#list>
      </#if>
      ... 
}
\end{lstlisting}
\end{figure*}

\subsubsection{Templates for SPIR-V Instructions}

Each SPIR-V instruction that was parsed from the initial JSON file will correspond to a new Java class that will be generated based on a template within the TSE component dedicated to SPIR-V instructions. 

Listing~\ref{code:instructionGenerator} shows a pseudo-code of the TSE component that generates all Java classes for all the SPIR-V instructions. 
Line 1 opens a file that represents the template in which the instructions will be written.
Then, lines 2-7 traverse the array that contains all parsed instructions from the input JSON file to generate a Java class per instruction. 
Each file contains a set of special characters represented as strings that the template system engine takes to substitute for specific values and new characters.
The end result is a set of valid Java classes that compose the whole SPIR-V Java library. 

Listing~\ref{code:templateInstruction} shows a code snippet of a Java template class for generating all instructions for the \toolkit. 
To achieve that, the TSE component makes use of the Apache Freemarker framework.
As shown in line 2 of Listing~\ref{code:templateInstruction}, all SPIR-V instructions extend a common Java class. 
The common Java class is stored in a hash table within the TSE software component that maintains a mapping between each of the SPIR-V instructions and the corresponding Java superclass, which was initially stored in the SPIR-V Library Generator. 
Each property from the Java class template corresponds to a field mapped from JSON parsed class (e.g., line 6). 

For SPIR-V instructions that contain optional operands, we generate the list of subsequent Java classes  by using special characters in the template that tells the generator to generate a set of field properties for each of the optional arguments, as we describe in line 4 of Listing~\ref{code:templateInstruction}. 
This syntax is provided by the Apache FreeMarker framework. 

As a result of the processing of all the SPIR-V instructions, the TSE component dynamically generates \textit{366 new Java classes} (all instructions available in the SPIR-V 1.2), and \textit{667 new Java classes} for the 1.6 version of the SPIR-V standard, while all instruction classes inherit from the same \texttt{SPIRVInstruction} base class that we provide.

\subsubsection{Templates for SPIR-V Operands}

Similarly to the templates for the instructions, the TSE component generates a set of Java types for all SPIR-V operands and SPIR-V kinds that will be then copied to a Java sub-package of the final library.
However, in this case, the operands are not generated just from a single template, but from three templates.

To comply with the SPIR-V SPEC, the SPIR-V TSE component provides a template for operands of \textit{composites}, \textit{enumerates}, and \textit{literals}. 
Apart from having more templates for the operands, the generation of the new Java classes for each operand is identical to the process of generating the corresponding classes for the instructions. 

In total, the TSE software component generates \textit{34 Java classes} for the SPIR-V 1.2 and \texttt{44 new Java classes} for the SPIR-V 1.6 standard, that are mapped to different types of SPIR-V operands, such as \texttt{SPIRVBuiltin}, \texttt{LiteralExtIns}-\texttt{Integer}, etc.
Note that this process is fully automatic, and it is triggered every time the \toolkit is built from the source code.

\subsubsection{Templates for Proxy Classes and Mappers}

With the generated Java classes that compose the Java Library for SPIR-V instructions and operands, it is possible to dynamically build SPIR-V modules just by using the generated composable and functional API. 
However, the SPIR-V library can be also used to disassemble SPIR-V binary modules as well as to assemble modules expressed in a text format. 
To achieve that, the TSE software component generates proxy classes that map both the description of the input SPIR-V module in a text format to binary (assembling); and the reverse action from binary modules to text (disassembling).

The TSE component generates three new classes per Java module: one module for the assembler, one module for the disassembler, and the Java classes that correspond to the mappers for instructions, operands, and extended OpenCL instructions (e.g., OpenCL math operations).
We define these proxy classes as Java helpers for writing SPIR-V and disassembling SPIR-V binary modules. 
As a result, once the SPIR-V Library is generated, it is ready to be consumed by client applications.

\subsection{SPIR-V Library}
\label{sub::library}

The result of the SPIRV Library Generator through its JSON parser and its TSE component is the SPIR-V Library, which contains an API along with a set of Java types to compose SPIR-V modules through instructions and operands.
The generated library also includes an interface as well as a set of mapper classes for the assembler and disassembler of SPIR-V modules. 

The generated API has two main Java types: i) the instructions and operands (as we discussed in Section~\ref{sec::templateSystem}); and ii) the scopes.
A scope data type represents a block type that defines different visibility regions within the SPIR-V module. 
There are three main types of scope data types in SPIR-V: a) modules; b) functions; and c) blocks (basic block of instructions), as shown in Figure~\ref{fig::apiBlocks}.
The outermost scope for a SPIR-V is a module and it holds the global properties (e.g.,\ header, variables, addressing modes, capabilities, etc.) that affect the whole module (all compute kernels that are contained in it).  

\subsubsection{Handling Different Scopes within a SPIR-V Module}
\begin{figure}[b!]
    \centering
    \includegraphics[width=\columnwidth]{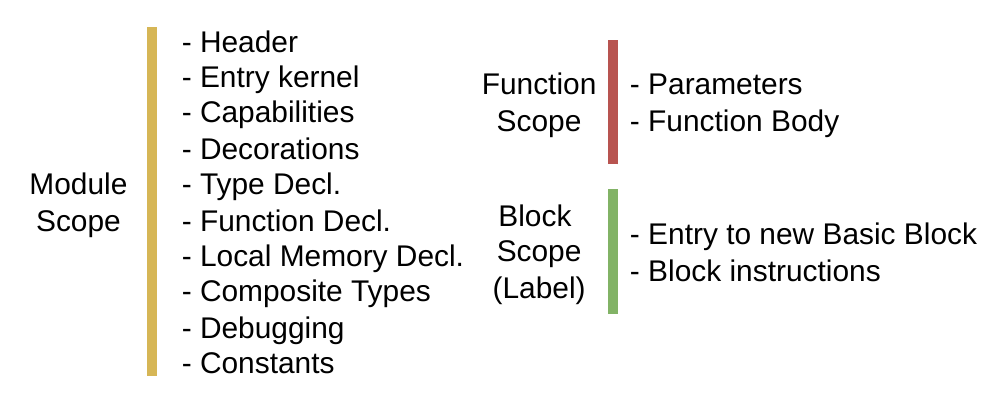}
    \caption{Different Scopes for a SPIR-V Module within the Beehive SPIR-V Toolkit API.}
    \label{fig::apiBlocks}
\end{figure}

Each type of scope in SPIR-V (i.e.,\ module, function, block) handles specific types of instructions. 
Figure~\ref{fig::apiBlocks} shows the three different scope objects that are available in the \toolkit. 

\paragraph{Module Scope}
The declaration of global properties (e.g.,\ SPIR-V capabilities, SPIR-V composite types, SPIR-V decorations, etc.) is controlled by the \textbf{module} scope. 
The instructions allowed in this scope are the header declaration, the capabilities for the module (e.g., a compute \textbf{kernel}, a shader, enabling \texttt{fp64}, etc), decorators (e.g., to specify memory alignment of variables), types and composite declaration and function declarations. 
The module scope also handles the declaration of the device's local memory (e.g., shared memory on a GPU device).

\paragraph{Function Scope}
Regarding the function scope, it handles the declaration of the parameters passed and declared at the function level and the function body. 

\paragraph{Block Scope}
Finally, the block scope handles the rest of the instructions of a basic block.
In this paper, a basic block corresponds to a sequence of instructions that are executed one after the other and it does not contain control flow divergence. 
As soon as there is a new basic block, a new SPIR-V label must be instantiated, and therefore, the new instruction will be enclosed in a new block scope.

Listing~\ref{code:exampleAPI} shows an example of how to build the header for a new SPIR-V module using the \toolkit API. 
A SPIR-V module receives as a parameter, a SPIR-V header object, which is also a Java type provided by the SPIR-V Library. 
The SPIR-V header arguments specified in the constructor of the object correspond to the parameters specified in the SPIR-V standard, exactly in the same order. 
In fact, we use the same order of the arguments as described in the SPIR-V standard.
Thus, this makes it easier to follow the SPEC along with the API definition of the \toolkit. 

As a side note, in the case of the argument for the initial bound (fourth argument from the \texttt{SPIRVHeader} object constructor), is set to zero. 
This is because the number of declared variables is unknown since this call is invoked at the initial stage of building the SPIR-V module. 
Therefore, the SPIR-V header is built with that value set to zero, and the library will recalculate that number right before writing the SPIR-V module into a file (or any other output via Java stream).

\begin{figure}[t!]
\begin{lstlisting}[caption=Example of how to build a header for a SPIR-V module., label={code:exampleAPI}, xleftmargin=.03\textwidth, language=Java] 
SPIRVModule module = new SPIRVModule( 
    new SPIRVHeader( 
      1,  // SPIRV_MAJOR_VERSION
      2,  // SPIRV_MINOR_VERSION
      32, // SPIRV_GENERATOR_ID
      0,  // SPIRV_INITIAL_BOUND
      0)); // SCHEMA_BOUND
\end{lstlisting}
\end{figure}

\subsubsection{Enabling Function Composition}

To enable the function composition of SPIR-V instructions, we designed a Java common interface (called \texttt{SPIRVInstScope}) for all types of SPIR-V scopes within the library.
This interface contains abstract methods for adding new instructions to a SPIR-V module and registering new identifiers (IDs) to the SPIR-V module. 

Listing~\ref{code:scopeInterface} shows a simplified view of the \texttt{SPIRVInstScope} Java interface that the Beehive SPIR-V Toolkit library provides. 
There are two main methods for this interface: one for adding instructions (\texttt{add}), and another one for registering new IDs within the SPIR-V module that is being built (via \texttt{getNextId}).

\paragraph{Handling SSA Variables}
In the \toolkit API, each instruction and kind have a unique SPIR-V identifier (ID).
The reason is that SPIR-V binaries follow the SSA (Static Single Assignment) representation~\cite{10.1145/73560.73562}, in which each SPIR-V construct is assigned once. 
Thus, the generated library provides the functionality to require and build unique IDs for any new instruction or kind declared in a SPIR-V module. 

To facilitate unique ID management, all new IDs are requested at the module level, which has the whole view of all instructions, operands, and kinds being used. 
Every time a new SPIR-V instruction is instantiated (a new Java object), the SPIR-V ID returned by the module is assigned to it (the new instruction). 
Additionally, since instructions are added within one of the three aforementioned scopes (module, function or block scopes), the moment that an instruction is added to the list of instructions that belong to the current scope, it also copies the ID of the instruction. 
In this way, the \toolkit library can perform checks on whether an ID is valid to be in the block that is being added. 

Listing~\ref{code:nextId} shows an example of how to obtain a new ID for the next instruction within a function scope level. 
In this case, we want to add a new SPIR-V label instruction. 
Thus, we request, at the module level, a unique ID. 
Recall that new IDs are requested from the module level, which has the global view of all IDs being used. 
This ID is then used to instantiate a new label instruction. 

The \texttt{add}, and \texttt{getNextId} methods in combination with all the instructions and operands generated by the TSE component, enable developers to compose SPIR-V modules and store them in binary format from Java. 

\begin{figure}[t!]
    \begin{lstlisting}[caption=Java Instruction Scope Interface to allow instruction composition., label={code:scopeInterface}, xleftmargin=.03\textwidth, language=Java] 
public interface SPIRVInstScope {
    // Add a new instruction
    SPIRVInstScope add(SPIRVInstruction var1);
    // Generate a new ID for the next instruction    
    SPIRVId getNextId();
}        
\end{lstlisting}
\end{figure}

\begin{figure}[t!]
    \begin{lstlisting}[caption=Example of obtaining new a new ID for the next instruction within a SPIR-V module., label={code:nextId}, xleftmargin=.03\textwidth, language=Java] 
SPIRVId idLabel = module.getNextId();
SPIRVOpLabel newLabel = new SPIRVOpLabel(idLabel);
functionScope.add(newLabel);
\end{lstlisting}
\end{figure}

\subsubsection{Examples}

This section shows a set of examples of how to invoke the \toolkit APIs for different case scenarios.

\paragraph{Generating SPIR-V Capabilities} 
Every SPIR-V binary module contains a set of capabilities. 
The capabilities define the memory addressing mode, types used, type of binary (graphics or compute), etc.
Capabilities are defined at the module level.

Listing~\ref{code:exampleCapabilities} shows a code snippet that adds the \texttt{Addresses}, \texttt{Linkage}, \texttt{Kernel}, \texttt{Int64}, and \texttt{Int8} capabilities into a SPIR-V module.
Since all the Java classes that represent instructions inherit from the same Java base class (\texttt{SPIRVInstruction} class), it facilitates the composition of different SPIR-V binary modules.

\begin{figure}[h!]
    \begin{lstlisting}[caption=Example of adding SPIR-V Capabilities to a SPIR-V Module for Compute., label={code:exampleCapabilities}, xleftmargin=.02\textwidth, language=Java] 
module.add(new SPIRVOpCapability(
           SPIRVCapability.Addresses()));     
module.add(new SPIRVOpCapability(
           SPIRVCapability.Linkage()));       
module.add(new SPIRVOpCapability(
           SPIRVCapability.Kernel()));      
module.add(new SPIRVOpCapability(
           SPIRVCapability.Int64()));     
module.add(new SPIRVOpCapability(
           SPIRVCapability.Int8())); 
\end{lstlisting}
\end{figure}

Listing~\ref{code:exampleCapabilitiesGenerated} shows the generated SPIR-V disassembled binary code in a text form that corresponds to the Listing~\ref{code:exampleCapabilities}.

\begin{figure}[h!]
    \begin{lstlisting}[caption=Disassembled SPIR-V code generated from Listing~\ref{code:exampleCapabilities}., label={code:exampleCapabilitiesGenerated}, xleftmargin=.02\textwidth, language=Java] 
OpCapability Addresses
OpCapability Linkage
OpCapability Kernel
OpCapability Int64
OpCapability Int8
\end{lstlisting}
\end{figure}

\paragraph{Addition} 
Listing~\ref{code:exampleAdd} shows an example to generate an integer addition instruction in SPIR-V (\texttt{OpIAdd}). 
Line 1 requests a new ID for the module. 
All IDs must be requested at the level of the SPIR-V module scope, which, again, is the scope type that has the whole view of the IDs for every declared variable. 
Then, line 2 creates an instance of the object \texttt{SPIRVOpIAdd} while passing the following parameters: a) the result type id (line 3) that must be previously declared in the module; b) the result ID (line 4); c) and the two operands IDs (lines 5-6).
In this case, the instructions are added using the block scope level.

\begin{figure}[t!]
    \begin{lstlisting}[caption=Example of an Addition using the Beehive SPIR-V Toolkit API., label={code:exampleAdd}, xleftmargin=.02\textwidth, language=Java] 
SPIRVId resultAdd = module.getNextId();
blockScope.add(new SPIRVOpIAdd(
    resultTypeId, 
    resultAdd, 
    operand1, 
    operand2));
\end{lstlisting}
\end{figure}

\paragraph{Expressing Control-Flow} 

\begin{figure*}[t!]
    \centering
    \includegraphics[width=0.8\textwidth]{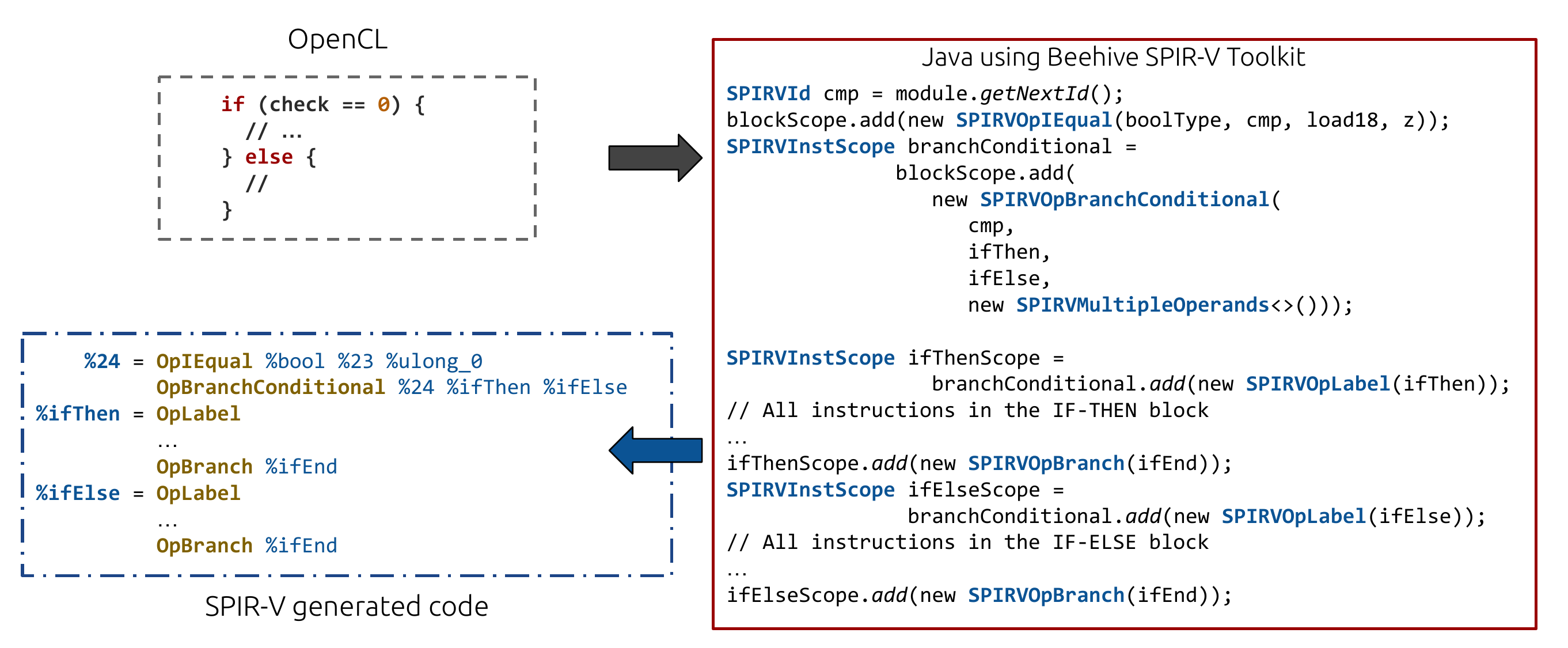}
    \caption{IF-ELSE composition of calls using the Beehive SPIR-V Toolkit API. The top-left box represents the equivalent OpenCL C. The right-hand side box shows a sketch of Java code using our API. The bottom right shows the disassembled generated code from the library.}
    \label{fig::apiBlocksConditions}
\end{figure*}

Figure~\ref{fig::apiBlocksConditions} illustrates a more complex example of how an OpenCL \texttt{if-else} conditional statement is programmed by using the \toolkit API for generating the equivalent code in SPIR-V.
Figure~\ref{fig::apiBlocksConditions}-top left box shows the OpenCL C code used in this example to generate SPIR-V code.
The box on the right-hand side shows the Java code that enables the \texttt{if-else} condition to be converted in SPIR-V instructions.
The box on the bottom-left represents the disassembled SPIR-V code that was generated through the \toolkit library. 

The first instruction from the Java code on the left corresponds to the \texttt{boolean} condition of the if-statement.
The next instruction is a conditional branch. 
To emit the conditional branch instruction, it is necessary to have all SPIR-V IDs registered already (\texttt{cmp}, \texttt{ifThen}, and \texttt{ifElse} IDs from the Java code example).
The next instruction that is generated is a new SPIR-V label for each block; one label (\texttt{\%ifThen} or \texttt{\%ifElse}), depending on the value of the condition.
Every time a new label is registered, it returns a new block scope. 
The new block scope must be used to compose all instructions that belong to the current scope.
For example, the Java code on the right-hand side of Figure~\ref{fig::apiBlocksConditions} shows the introduction of a new block called \texttt{ifThenScope}.
That block is used to generate all instructions in the \texttt{\%ifThen} basic block. 

\subsubsection{Dis/Assembler}

As we introduced in Section~\ref{sec::templateSystem}, the TSE component also generates two Java packages with all the logic regarding the assembler and disassembler. 
These two packages mainly contain the Proxy (or mapper) Java classes to transform from a) a text file that describes a SPIR-V module to a SPIR-V binary (assembler); and b) from a SPIR-V binary to a text file. 
Note that, if developers want to generate new SPIR-V modules in an instruction-by-instruction manner, the use of these mapper classes is not needed. 
However, these mapper Java classes are required in order to assemble or disassemble a SPIR-V binary code (from SPIR-V text to binary and vice-versa). 

In order to generate the correct SPIR-V binary file from a text file that describes a SPIR-V module, the mapper Java classes contain helper methods.
These methods are used to transform each token (i.e.,\ instruction from the input file) to a SPIR-V instruction class by composing instructions of modules, functions and blocks (as explained in the previous section).
Additionally, the Java mapper classes for the assembler keep a mapping between all unique IDs from the input text file in order to be used for the instructions that require these IDs as operands. 
For example, when a new variable is declared, the mapper Java class creates a mapping between the assigned string name and a new \texttt{SPIRVId} object that is created, and it is stored in an internal hash table.
Then, when the requested ID is passed as an argument to any SPIR-V operation, the mapper recovers the \texttt{SPIRVId} by performing a lookup in the hash table. 

Note that the input text file must represent a valid SPIR-V code. 
This is similar to the \texttt{spirv-as} LLVM utility command from the Khronos SPIR-V utilities~\cite{spirvtools}, 
in which the SPIR-V text file can be manipulated and tested before integrating optimizations in the compiler pipelines of the optimizing compilers, thereby facilitating fast debugging. 
Since SPIR-V is an intermediate representation in binary format, it is not easy to try out new optimizations and reordering operations without changing and adapting the compiler infrastructure. 
However, the implemented approach allows developers of JIT compilers to test new optimizations, measure performance and then integrate the changes in the compilation pipeline while using the Java software ecosystem. 

The disassembler mapper classes work in a similar way to the assembler mapper classes but in the reverse order. 
The mapping classes contain the logic to transform SPIR-V opcodes (integer values) into text format. 

\begin{figure}[t]
    \begin{lstlisting}[caption=Code snippet that shows how to invoke the disassembler., label={code:dis}, xleftmargin=.02\textwidth, language=Java] 
SPIRVDisassemblerOptions opt = 
    new SPIRVDisassemblerOptions(
      true,  // Syntax Highlight
      true,  // Show Inline Names
      false, // Turn Off Indentation
      true,  // Should Group
      false); // No Header
SPIRVTool spirvTool = new Disassembler(
      reader,     // input binary file
      System.out, // output 
      opt);       // options
spirvTool.run();
\end{lstlisting}
\end{figure}

Listing~\ref{code:dis} shows an example of how to disassemble a SPIR-V binary file into text. 
Line 1 creates a Java object for selecting the options for the disassembler. 
The constructor accepts some utilities that can be configured, such as syntax highlighting, indentation, turning off the header, etc. 
Lines 8-11 create the SPIR-V disassembler object and line 12 invokes the runner. 
When lines 8-11 are executed, it prints the SPIR-V text file that corresponds to the input binary in the selected output (e.g.,\ standard output).

\begin{figure*}[t!]
    \begin{lstlisting}[caption=Example of how to invoke the client application for running the assembler and disassembler., label={code:client}, xleftmargin=.04\textwidth, language=Bash] 
## Run the client application to transform a SPIR-V binary to a text file
$ java -jar dist/spirv-beehive-toolkit.jar -d kernel.spv -o output.spirvText
## Run the client to transform a SPIRV text file into a SPIR-V binary
$ java -jar dist/spirv-beehive-toolkit.jar -d  --tool asm -o finalBinary.spv output.spirvText
\end{lstlisting} 
\end{figure*}

\subsubsection{Validation Rules}
The \toolkit also provides some validation methods for the generated SPIR-V binary modules. 
The validation rules provided are as follows:
the SPIR-V module must have at least one function declared, at least one capability declared, the memory model must be defined and it should contain at least one entry point. 
Furthermore, the generated SPIR-V binary modules can be validated using the Khronos SPIR-V Tools, such as \texttt{spirv-val}.
The validation of the generated modules is important to ensure the fidelity of the code generator.
Our validator is complementary to the SPIR-V Khronos validator. 
For instance, our validator also checks for instruction capabilities and dependencies before storing the final SPIR-V binary.
The dependencies are retrieved from the JSON file that describes the SPIR-V grammar.
Through this validation, we were able to detect circular dependencies in the latest SPIR-V standard (1.6).

\subsubsection{Properties of the \toolkit Library}
The key properties of the Beehive SPIR-V Toolkit library are:

\paragraph{Compliance:} The generated library stores all the instructions and operands in the same format as specified in the standard. Each instruction is 32-bit wide and contains all operands and opcodes. 
Additionally, the generated binaries by the library can be validated using the Khronos SPIR-V Tools, such as \texttt{spirv-val}.

\paragraph{Composability \& Maintenance:} Developers can build SPIR-V modules in a composable and functional style when using the generated API of the SPIR-V Library. 
We believe that this way of programming SPIR-V modules from high-level programming languages and interfaces such as Java and JVM languages makes the code easier to understand, maintain, and debug.

\paragraph{Integration with JVM languages:} Since the Beehive SPIR-V Toolkit is fully implemented in Java, it can be integrated into any Java framework and it can be invoked by any JVM programming language (e.g., Truffle framework~\cite{10.1145/3062341.3062381, 10.1145/2509578.2509581}). 
For example, it can be invoked by state-of-the-art compiler frameworks for JVM such as the Truffle framework~\cite{10.1145/3062341.3062381, 10.1145/2509578.2509581}, which implements programming languages, such as R~\cite{10.1145/3093334.2989236}, JavaScript~\cite{10.1145/3062341.3062381}, or Ruby~\cite{10.1145/3237009.3237026}, among many others. 

Since all these languages are implemented in Java and on top of the JVM, it is easier for those programming languages to be able to generate SPIR-V kernels (either for enabling computation or for graphics processing).
Note that this is not currently implemented, but, since the proposed library is open-sourced, the JVM and Truffle community could benefit from this library. 

\paragraph{Rich Ecosystem and Tooling:} Java has a rich programming ecosystem, containing multiple IDEs (such as Eclipse, NetBeans or IntelliJ), debuggers, and profilers.

\subsection{SPIR-V Client Utility}
\label{sub:client}

The \toolkit also contains a standalone module that includes a client application for assembling and disassembling SPIR-V code. 
The client application acts as a command line utility that invokes the assembler and disassembler components from the library. 

Listing~\ref{code:client} shows an example of how to invoke the client utility to disassemble and assemble SPIR-V kernels. 
Line 2 invokes the disassembler that takes as input a SPIR-V kernel (named ``kernel.spv'' that can be compiled, for instance, using LLVM/CLANG), and it generates the description of the disassembled code in a text file. 
Line 5 performs the reverse process.
In essence, it takes as input a file that contains the description of the SPIR-V code in text format, and it generates a SPIR-V binary (named ``finalBinary.spv''). 
Furthermore, between line 2 and line 5, developers and users can modify the SPIR-V code in the text format and generate a new binary from it.

\section{Use case: Integration into TornadoVM}
\label{sec::usecases}

\begin{figure*}[t!]
    \centering    
    \includegraphics[width=1\textwidth]{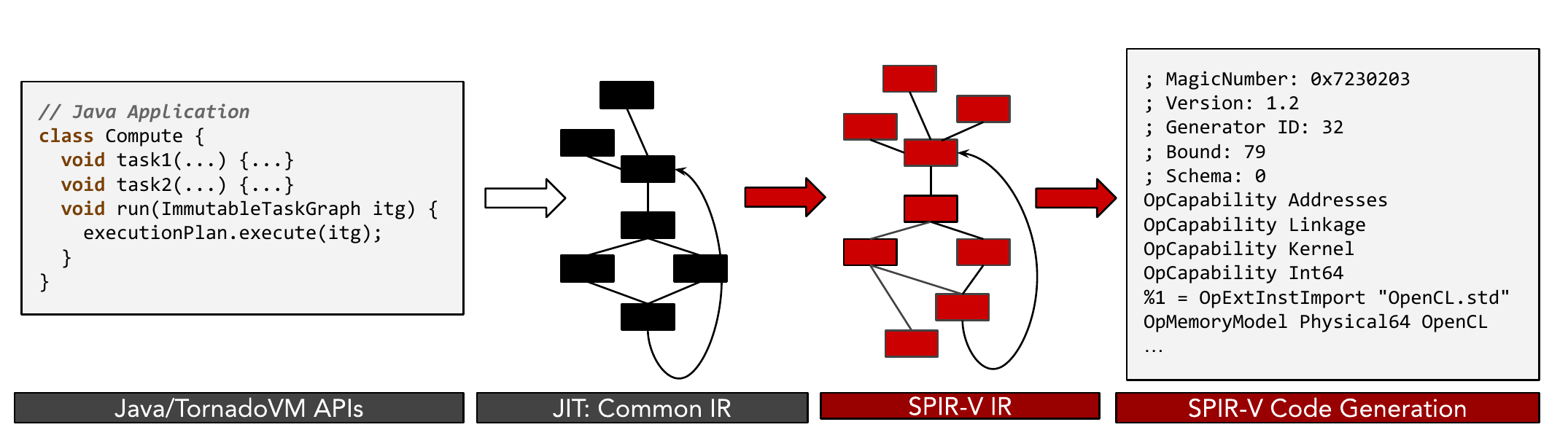}
    \caption{SPIR-V JIT Compilation Process in TornadoVM.}
    \label{fig::tornadovm}
\end{figure*}

Once we developed the library generator, we integrated it into the TornadoVM JIT compiler. 
TornadoVM~\cite{10.1145/3313808.3313819, 10.1145/3237009.3237016} is a Java parallel programming framework that transforms, at runtime, Java bytecode into OpenCL and PTX code. 
To do so, TornadoVM accelerates a subset of Java by offloading code to be executed on CPUs, GPUs and FPGAs. 

We extended the TornadoVM project with a new backend for dispatching SPIR-V code through the Level Zero~\cite{levelZero} API (a new low-level API developed by Intel to manage heterogeneous devices). 
Although the Level Zero API is not attached to any specific hardware accelerator, it is currently only available for Intel-integrated GPUs. 
Thus, we prototype the SPIR-V backend for integrated GPUs with support for Level Zero and SPIR-V. 
In the future, we plan to also integrate the SPIR-V backend to be also dispatched through the OpenCL runtime.

\paragraph{SPIR-V Compilation Process for TornadoVM} 
Since the TornadoVM JIT compiler extends the Graal JIT compiler~\cite{duboscq2013graal}, and it is fully implemented in Java, we imported our library as a new dependency for the SPIR-V backend within TornadoVM. 
We implemented the new backend following the guidelines for the existing backends in TornadoVM such as the OpenCL C and the NVIDIA PTX backends. 

The compilation workflow for the SPIR-V backend is shown in Figure~\ref{fig::tornadovm}.
The Figure is divided into four blocks: the first block on the left shows a Java code snippet that represents a program using the TornadoVM parallel APIs for GPU and FPGA programming. 
When the code is first executed, the TornadoVM runtime system invokes the JIT compiler for each Java method to be compiled to the target backend (e.g., SPIR-V). 

TornadoVM lowers the code from the Java bytecode to the target backend in different phases or tiers. 
The first tier is called the sketches, and it contains a common representation for all backends (e.g., OpenCL and SPIR-V).
During the sketches, the TornadoVM JIT compiler applies common high-level optimizations, such as constant folding, evaluation of expressions, etc. 
We did not extend this phase to adopt the new SPIR-V Backend. 

From the sketcher, the TornadoVM JIT compiler starts specializing the IR per backend. 
In our case, we extended with a new set of compiler optimizations and lowering phases to transform the high-tier compiler intermediate representation of the program into the SPIR-V code. 
To do so, we followed the compilation pipeline of TornadoVM with three new compilation tiers, named high-tier, mid-tier and low-tier, and added, in total, 58 new compilation phases that are specialized for the SPIR-V backend.
These compilation phases include optimizations for performing fast math operations, vector operations, and inlining, among many others. 

Once the TornadoVM JIT compiler optimizes the IR, we generate the corresponding SPIR-V code using our \toolkit library. 
To generate SPIR-V, we traverse the final IR and build a new list of \textit{lowerable} Java objects (Java types provided by GraalVM to generate code) with the specific instructions to generate SPIR-V code. 
The final process is to traverse the final list and invoke the \texttt{generate} method for each object in the list. 
The result of this process is a SPIR-V module that can be dispatched through the Level Zero API included in TornadoVM. 

\paragraph{Advantages of the SPIR-V Backend in TornadoVM}
From our experience porting the SPIR-V backend in TornadoVM, we see that the structure is much simpler than the OpenCL C backend.
This is because the SPIR-V backend generates code at the binary level, while the OpenCL C code reconstructs source code from the low-level IR of the TornadoVM JIT compiler. 

For instance, the OpenCL C backend contains a lot of control to cover many corner cases when generating structured control flow from the unstructured control flow of the TornadoVM IR, and therefore, the Graal IR~\cite{10.1145/2936313.2816715}.
The SPIR-V backend of TornadoVM dramatically simplifies this process by allowing conditional and unconditional jumps to the specific compiler basic blocks. 

The \toolkit library can also be used in an analogous manner by other runtime systems and optimizing JIT compilers, such as FastR-GPU~\cite{10.1145/3050748.3050761}, Marawacc~\cite{Fumero17:AIPGPUs}, Aparapi~\cite{aparapi.github.io}, or IBM GPU J9~\cite{ibm-j9-2015}, Python MegaGuards~\cite{qunaibit_et_al:LIPIcs:2018:9221} or Ruby GPU~\cite{10.1145/2935323.2935327}.

\section{Evaluation}

\begin{figure*}[t]
    \centering    
    \includegraphics[width=0.9\textwidth]{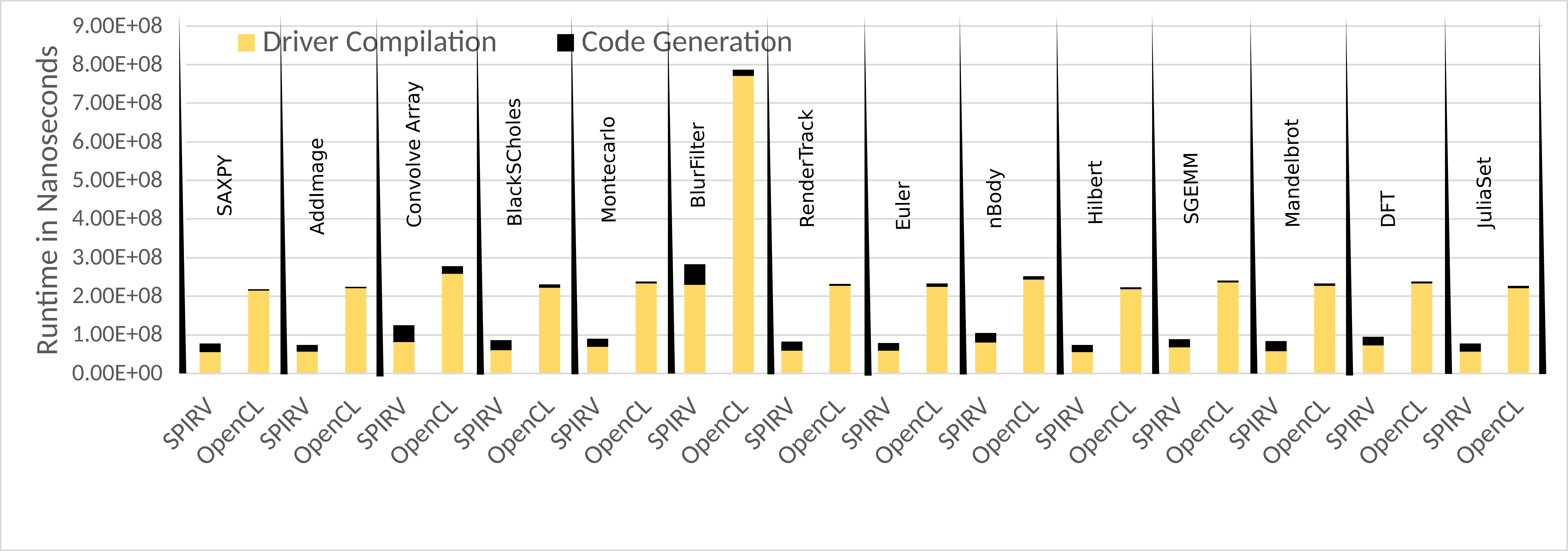}
    \caption{Performance evaluation between the code generator of TornadoVM for OpenCL and the SPIR-V. The SPIR-V backend uses the Beehive SPIR-V Toolkit library. 
    The stack plot displays two types of metrics: a) the driver compilation, which involves the JIT compilation once the code is generated; and b) the code generation, which contains the total time to generate the code by using the proposed library.
    The total time is shown in nanoseconds. The lower, the better.}
    \label{fig::evaluation}
\end{figure*}

We evaluated the library against the existing OpenCL Backend in TornadoVM running on the same device, and Intel-integrated GPU.
\paragraph{Setup} We use an Intel HD Graphics GPU contained in an Intel i9-10885H CPU. 
The Intel driver used was 21.38.21026, which is the same for running OpenCL and SPIR-V applications.
The OS used was Fedora Linux 34 with the Linux Kernel 5.16.18-100.
TornadoVM contains a script for benchmarking, and it includes applications from many different domains such as machine learning, Fintech, linear algebra and physics~\cite{10.1145/3313808.3313819}.

\paragraph{Performance of the JIT Compiler and Code Generation}
Figure~\ref{fig::evaluation} shows the execution time for code generation (the total time that takes to build either a SPIR-V module or an OpenCL C kernel from the last compiler phase), and the driver compilation, which represents the total time that the GPU driver takes to compile to the final GPU binary. 
The SPIR-V backend employs the \toolkit library, while the OpenCL C backend of TornadoVM uses an ad-hoc library of the project (with no decoupling).

In general, the SPIR-V backend performs slower than the OpenCL when generating the code (black section of the stacks in Figure~\ref{fig::evaluation}).  
This is expected since the \toolkit API creates a significant number of Java objects in order to build SPIR-V modules.
This is because each identified and each instruction is a new Java object. 
However, once the code has been generated, the driver compilation time decreases by up to 3.9x compared to OpenCL. 

Although the function composition and modularity of the SPIR-V library come at a performance cost, the most efficient driver implementation of SPIR-V results in an end-to-end compilation performance increase of 2.72x (on average) compared to OpenCL. 
End-to-end performance speedups, compared to OpenCL, range from 2.2x\%{} (\textit{convolveImageArray} - the third group of bars), to 3x (\textit{hilbert} - the fifth group of bars starting from the right-hand side).

\paragraph{Performance of the SPIR-V Backend}

\begin{figure}[t!]
    \centering    
    \includegraphics[width=1\columnwidth]{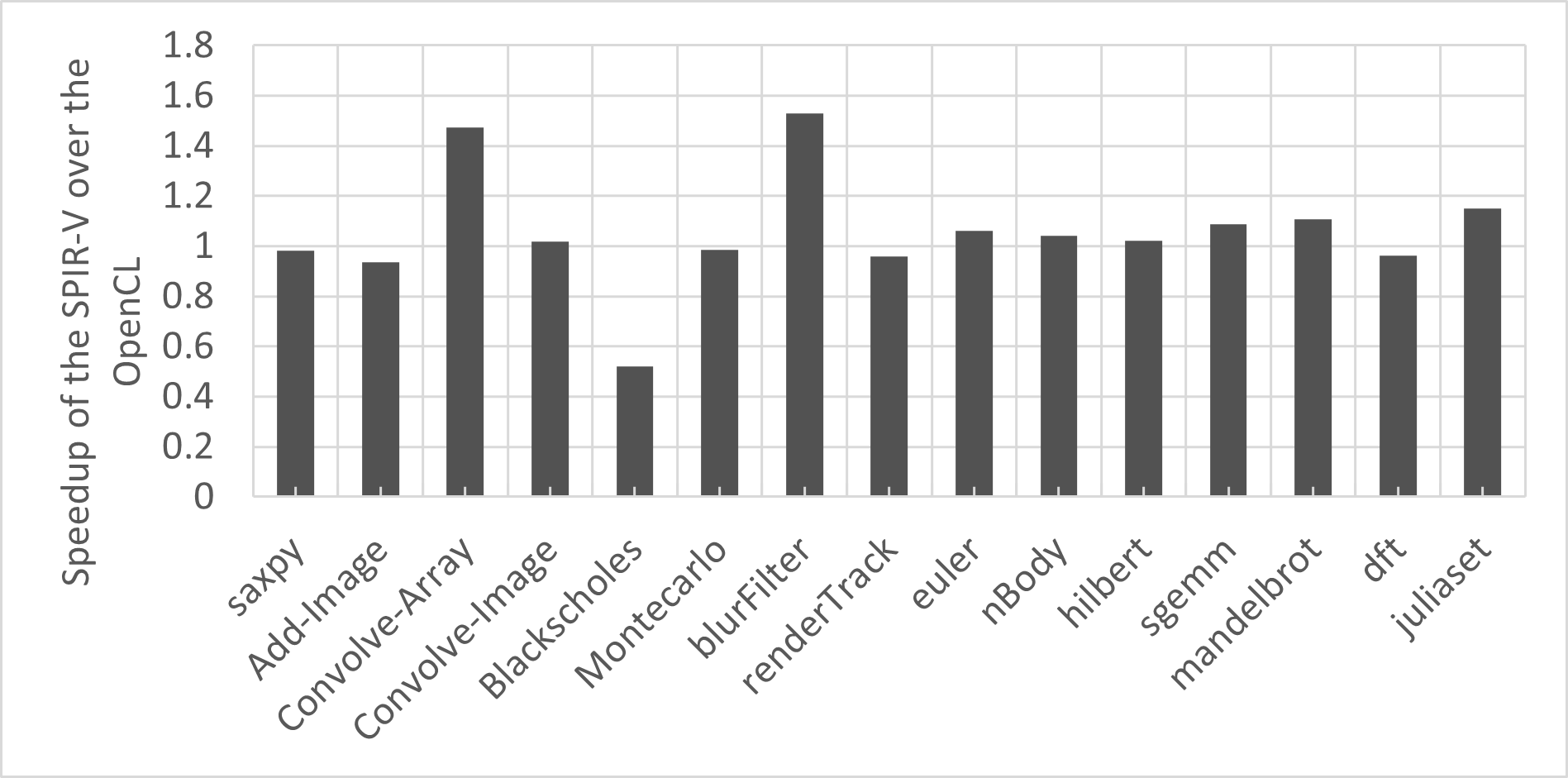}
    \caption{Speedup of the our implementation of the SPIR-V backend for TornadoVM using the \toolkit library API over the existing OpenCL C backend on the integrated Intel GPU.}
    \label{fig::evaluationTotal}
\end{figure}

We also evaluated the total time that takes each benchmark to run for each backend. 
Figure~\ref{fig::evaluationTotal} shows the speedup of each application using our new SPIR-V backend over the OpenCL C backend that was already implemented in the TornadoVM project. 
Thus, the higher, the better. 
We executed all benchmarks on the same Intel HD graphics for both backends. 

To measure the performance, we executed the benchmark script that the TornadoVM provides, which performs a set of warm-up iterations and then provides the median time for all executions.
Since we provide the median, the JIT compilation time, which happens in the first iteration, is excluded from these measurements. 

As a disclaimer, note that the SPIR-V backend is dispatched through the Intel Level Zero API, while the OpenCL C backend is dispatched through the Intel OpenCL driver for the same GPU (Intel integrated GPU). 

We see that, in general, the SPIR-V backend varies from 2\%{} slowdown compared to the OpenCL C backend execution (for the \texttt{saxpy}, \texttt{addImage} and \texttt{dft} benchmarks) to 15\%{} speedup (for the \texttt{juliaSet} benchmark.

In particular, the SPIR-V performs much better than OpenCL for the \texttt{convolveArray} and \texttt{blurFilter} benchmarks, 1.47x and 1.52x respectively. 
In contrast, the \texttt{blackScholes} benchmark performs 50\%{} slowdown compared to the OpenCL C. 

Even though we are executing the benchmarks on the same device, we see different performances.
One of the reasons is that the thread block we select for the SPIR-V backend is different compared to the one that TornadoVM selects for the OpenCL C backend.
This clearly can influence performance~\cite{10.1007/s11227-013-0921-z}. 

\section{Related Work}
\label{section:relatedwork}


\paragraph{Template System Engine}
SPIR-V Tools~\cite{spirvtools} uses a similar approach for mapping the code between SPIR-V text and SPIR-V binary. 
However, our approach also provides a functional programming interface to dynamically build SPIR-V modules embedded in the template system engine.
We believe this API offers a cleaner and more concise way to compose SPIR-V applications. 

LLVM \textit{TableGen}~\cite{llvm} also uses a similar approach to our template system engine.
In LLVM, \textit{TableGen} is a tool provided by the LLVM compiler infrastructure that can be used to generate code using the \texttt{llvm-tblgen} utility command.
LLVM \textit{TableGen} is a more powerful tool than our template system engine because it can be used not only to generate operands and instructions (as the Beehive SPIR-V Toolkit does), but also it can generate instruction scheduling information and specific rules for each instruction. 
However, our template system engine can be extended to accommodate this kind of functionality by providing the corresponding templates in Java.

J. Haavisto~\cite{haavisto:hal-03155647} presented an approach to generate SPIR-V binary modules using programs written in APL~\cite{10.5555/1098666} (a non-imperative and array centric programming language) as a modelling language. 
There are two differentiation points with our work: 
i) our work exposes a generic and composable API, rather than the templates itself generate the whole SPIR-V code, 
and ii) the template system engine is generic to support other code instruction sets and intermediate representations such as OpenCL or CUDA PTX.

\paragraph{SPIR-V ASM/DASM libraries}
The Multi-Level IR Compiler Framework (MLIR)~\cite{DBLP:journals/corr/abs-2002-11054} contains a custom implementation of a SPIR-V library to generate SPIR-V binary modules from the MLIR-specific intermediate representation.
The custom SPIR-V library generates a dialect of SPIR-V (LLVM IR with SPIR-V intrinsics)\footnote{\url{https://groups.google.com/g/llvm-dev/c/n0vU71iHNis}}.
MLIR provides utilities to lower the SPIRV-V dialect to standard SPIR-V modules. 
The SPIR-V dialect generated by the MLIR is designed to perform specific optimizations~\cite{Vasilache2022ComposableAM}.
In contrast, the prime focus of the proposed template system engine is to generate standard SPIR-V. 
However, the set of templates in Beehive SPIR-V Toolkit can be extended to include the emission of other dialects that can also interact with the MLIR GPU dialects for SPIR-V.
To do so, we would need to extend the \texttt{write} methods of each SPIR-V instruction to emit the equivalent SPIR-V dialect instructions.

ViennaCL++~\cite{10.1145/3204919.3207894} 
can generate SPIR-V binary modules by invoking the LLVM SPIR-V compiler for input OpenCL kernels
The generated library of the Beehive SPIR-V Toolkit can generate SPIR-V binary modules via a composable and functional Java API. 
DCompute~\cite{10.1145/3204919.3204922} is a SPIR-V code generator library that reuses the LLVM-based D Compiler (LDC)~\cite{llvm} to statically compile programs written in the D language to SPIR-V. 
The Beehive SPIR-V Toolkit, although it provides a Java API, it is language agnostic, making it suitable for reusing with other JVM languages such as Scala, R~\cite{10.1145/2989225.2989236}, JavaScript and NodeJS~\cite{10.1145/3062341.3062381}, Python or Ruby~\cite{10.1145/3237009.3237026}. 
Furthermore, DCompute can compile high-level language constructs, such as lambda expressions. 
Unlikely, the proposed library is designed to operate at a lower-level, and it is intended for compiler engineers who need to generate and debug SPIR-V code from managed runtime programming languages. 

The vast majority of the SPIR-V compilers use LLVM tools~\cite{spirvtools} to generate, analyze and validate the generated code.
Several compilers that belong to this category are the Intel SYCL compiler~\cite{10.1145/3318170.3318194}, HIPCL~\cite{10.1145/3388333.3388641}, ComputeCpp~\cite{computecpp, codeplayComputecpp}, HipSYCL~\cite{10.1145/3388333.3388658}, triSYCL~\cite{10.1145/3204919.3204937} and Intel oneAPI~\cite{oneAPI}.
Our approach leverages the SPIR-V code generation as a high-level library for Java and JVM-based programming languages.
\section{Conclusions}
\label{section:conclusions}

SPIR-V is an intermediate representation in a binary format to express parallel computations and graphics for execution on heterogeneous hardware.
SPIR-V tools that assemble and disassemble from/to parallel programming and graphics models such as OpenCL and Vulkan already exist. 
However, they are primarily addressed to programming languages, such as C/C++, and imported using the LLVM compiler infrastructure and software ecosystem. 

This paper presents the \toolkit, a Java programming framework that enables the generation of SPIR-V kernels from programming languages that can run on top of the JVM (e.g., Java, Truffle Python, Truffle Ruby or JavaScript).
While the technique to automatically generate programs from JSON files is not new, it has been exploited to generate a complete API from a specification, allowing the adoption of new releases of the standards with minimal effort quickly. 
The \toolkit is an open-source project that generates a functional and composable API for building SPIR-V modules. 

This paper also described the overall architecture of the proposed library and presented a template system engine that can automatically generate a programming library from a well-specified grammar written in JSON.
Furthermore, The \toolkit contains a client utility that can be used as a standalone tool for assembling, disassembling and validating a SPIR-V binary code.

We showcase the integration of our library into the TornadoVM project, by providing a new backend to compile Java to SPIR-V.
Our experiments show that our library performs up to 3x compared the existing OpenCL C code generator in TornadoVM, 
and the generated SPIR-V code performs up to 1.52x faster than the OpenCL C backend when running on an Intel integrated GPU.

\paragraph{Future Work}
In future work, we plan to extend validation rules.
Additionally, we want to expose an API for checking and validating different sections of the generated SPIR-V binary modules.
Furthermore, we aim to add support for further SPIR-V extensions from different SPIR-V vendors.  
Regarding the integration with TornadoVM, we plan to extend the the SPIR-V backend of TornadoVM to be used by its OpenCL runtime.


\begin{acks}
This work is partially funded by grants from Intel Corporation and the European Union’s Horizon 2020 programme under grant agreement No 957286 (ELEGANT).
Additionally, it is funded by UK Research and Innovation (UKRI) under the UK government’s Horizon Europe funding guarantee for grant numbers 10048318 (AERO), 10048316 (INCODE), 10039809 (ENCRYPT) and 10039107 (TANGO)
\end{acks}  

\balance

\bibliographystyle{ACM-Reference-Format}
\bibliography{main}

\end{document}